\documentclass[a4paper,twoside,openright,11pt]{article}

\pdfoutput=1

\usepackage[pdftex]{graphicx} 
\usepackage{breakurl}
\usepackage{enumerate}        
\usepackage{array}            
\usepackage{pdfpages}         
\usepackage{parskip}          
\usepackage{xcolor}           
\usepackage{makeidx}          
\usepackage{multicol}         
\usepackage{listings}
\usepackage{amsmath}
\usepackage{amssymb}
\usepackage{wrapfig}
\usepackage{multirow}
\usepackage{appendix}
\usepackage{marvosym}
\usepackage{longtable}
\usepackage{textcomp}

\usepackage{makeidx}
\RequirePackage[bookmarks,
                urlcolor=black,
                citecolor=black,
                linkcolor=black,
                pagecolor=black,
                colorlinks,
                hyperfigures]{hyperref}

\usepackage[utf8]{inputenc}

\addtocontents{toc}{\protect\thispagestyle{empty}}

\newcommand{\fitem}[1]{\item[]\hspace{-0.17in}\textbf{#1}}

\newcommand\minibox[2][l]{%
  \begin{tabular}{@{}#1@{}}
    #2
  \end{tabular}%
}

\let\oldmarginpar\marginpar
\renewcommand\marginpar[1]{\-\oldmarginpar[\raggedleft\footnotesize #1]%
{\raggedright\footnotesize #1}}

\newcommand{\rateA}{\marginpar{\Heart\Heart}}
\newcommand{\rateB}{\marginpar{\Heart}}
\newcommand{\rateC}{\marginpar{\textasteriskcentered}}
\newcommand{\rateD}{\marginpar{\texttimes}}

\definecolor{darkred}{rgb}{0.7,0.0,0.0}

\title{Cloud Process Execution Engine - Evaluation of the Core Concepts}
\author{J. Mangler$^{1}$ and G. Stuermer$^{2}$ and E. Schikuta$^{1}$\\
  \vspace{5mm}~\\
  University of Vienna, Faculty of Computer Science\\
  Workflow Systems and Technologies Group\\
  \vspace{5mm}~\\
  $^{1}$ \textit{juergen.mangler@univie.ac.at}\\
  $^{2}$ \textit{gerhard.stuermer@gmail.com}\\
  $^{3}$ \textit{erich.schikuta@univie.ac.at}
}

\begin{document}
  \maketitle

  \section{Introduction}
\label{sec:introduction}

In this technical report we describe describe the Domain Specific Language
(DSL) of the Workflow Execution Execution (WEE) introduced in
\cite{stuermer_domain_2009,stuermer_network_2009}. Instead of interpreting an
XML based workflow description language like BPEL, the WEE uses a minimized but
expressive set of statements that runs directly on to of a virtual machine that
supports the Ruby language. Frameworks/Virtual Machines supporting supporting
this language include Java, .NET and there exists also a standalone Virtual
Machine.

Using a DSL gives us the advantage of maintaining a very compact code base of
under 400 lines of code, as the host programming language implements all the
concepts like parallelism, threads, checking for syntactic correctness. The
implementation just hooks into existing statements to keep track of the
workflow and deliver information about current existing context variables and
state to the environment that embeds WEE.

  \section{Domain Specific Language}
\label{sec:dsl}

The DSL to describe a workflow consists of a minimal vocabulary. In this
section we want to describe the elements of the vocabulary in conjunction with
a simple example (see Listing \ref{lst:example}): a flight as well as a hotel
have to be booked for three people. If the payed sum for all three people
exceeds 10000 credits an external entity is informed. The set of available
DSL-elements is:

\begin{itemize}

  \fitem{activity} is an atomic operation and executes a specific task. We
  distinguish between two types of activities: \textit{manipulate-activities}
  (see line \ref{element:activity_manipulate}) are simple operations that are
  executed within the WEE. The intention is to provide an easy way to perform
  calculations or context changes. \textit{call-activities} (see line
  \ref{element:activity_call}) are used to carry out more complex tasks which
  are encapsulated in any kind of service. The execution of a
  \textit{call-activity} is delegated to the  handler wrapper. Therefore it is
  irrelevant if the service is a webservice or any other kind of service. The
  handler wrapper is provided with the location of the service (i.e. an
  endpoint) and optional parameters.

  \fitem{parallel} defines two or more parallel branches which are executed
  concurrently (see line \ref{element:parallel}). Each branch is executed in
  a separated thread but they still share the context of the workflow. We
  distinguish between two variants of parallel execution. First, the
  \textit{wait}-variant is on hold until each branch has finished before the
  thread of control is passed to the subsequent branch. Second, the
  \textit{nowait}-variant waits for a given amount of branches to be
  merged. Other branches who are not finished are informed to stop execution
  and are no longer executed by the WEE. The \textit{nowait}-variant
  implements a kind of race between the branches to finish. It can be used to
  start different approaches to finish a job, but continues as soon as any
  result is available.

  \fitem{choose} defines a decision in the control flow (see line
  \ref{element:choose}). Multiple or none of the available
  \textit{alternative}'s (guarded by a condition as shown in line
  \ref{element:alternative}) can be chosen.  Also, an else-path is provided
  by the \textit{otherwise}-keyword

  \fitem{cycle} enables top-controlled loops (see line \ref{element:cycle}). 

  \fitem{critical} implements the critical section pattern (see line
  \ref{element:critical}). A codeblock encapsulated in the critical-keyword is
  defined to be protected by a semaphore. In a multi-threaded environment, a
  critical section ensures exclusive execution. If a thread enters execution of
  a critical section, all other threads that want to enter this section have to
  wait until the first thread has exited the critical section.  Each critical
  section is defined by a symbolic name. Forming multiple codeblocks
  labeled with the same symbolic name enables critical sections to span over
  different parts of a workflow.
  
\end{itemize}

In addition to the control structures we also defined keywords for special
workflow purposes. These keywords are:

\begin{itemize}

  \fitem{handler} defines which handler wrapper should be included (see line
  \ref{element:handler}).

  \fitem{endpoint} defines the location of a service, i.e. an URI of a
  webservice (see line \ref{element:endpoint}). Endpoints are provided to the
  handler wrapper if an external service has to be invoked.

  \fitem{context} defines context variables of a workflow (see line
  \ref{element:context}). Each of the context variables is supervised by the
  execution engine. 

\end{itemize}

\lstset{language=Ruby,caption=\textsc{Example Workflow},label=lst:example}
\begin{lstlisting}
class Workflow < Wee
 handler MyHandler  (*@\label{element:handler}@*)
 
 endpoint :epAirBook => "uri:air/booking" (*@\label{element:endpoint}@*)
 endpoint :epHotelBook => "uri:hotel/booking"
 endpoint :epAirPay => "uri:air/payment"
 endpoint :epHotelPay => "uri:hotel/payment"
 endpoint :epInform => "uri:company/inform"

 context :persons => 3 (*@\label{element:context}@*)
 context :creditcard => "Visa_12345"
 context :airline => nil, :hotel => nil
 context :from => "Vienna", :to => "Prag"
 context :sum => 0

 control flow do
  cycle(@persons > 0) do (*@\label{element:cycle}@*)
   parallel(:wait) do (*@\label{element:parallel}@*)
    parallel_branch do
     critical(:airbooking) do (*@\label{element:critical}@*)
      activity :bookFlight, :call, epAirBook, @from, @to do |id| (*@\label{element:activity_call}@*)
       @airline = id
      end
      activity :payFlight, :call, epAirPay, @airline, @creditcard do |amount|
       @sum += amount
      end
     end
    end
    parallel_branch do
     critical(:hotelbooking) do
      activity :bookHotel, :call, epHotelBook, @to do |id|
       @hotel = id
      end
      activity :payHotel, :call, epHotelPay, @hotel, @creditcard do |amount|
       @sum += amount
      end
     end
    end
   end
   activity :countdown, :manipulate do (*@\label{element:activity_manipulate}@*)
    @persons -= 1
   end
  end
  choose do (*@\label{element:choose}@*)
   alternative(@sum > 10000) do (*@\label{element:alternative}@*)
    activity :inform, :call, epInform, @sum
   end
  end
 end
end
\end{lstlisting}

  \section{Evaluation}
\label{sec:evaluation}

In this section we try to evaluate the coverage of possible workflow control
structures with our basic set of DSL elements. We choose the workflow patterns
repository by Van der Aalst et al. \cite{van_der_aalst_workflow_2003} as
starting point to validate our prototype. The control flow patterns focus on
the dependencies between multiple activities and branches of a workflow. 

Each pattern can be supported by our execution engine in one of four different
ways which are ordered by degree of support:

\begin{itemize}

  \fitem{\rateA directly supported:} The pattern is supported by an explicit language
  element in the WEE-DSL or a simple combination of them.

  \fitem{\rateB modified workflow:} The pattern can be expressed by rearranging
  existing elements. This is possible if the behaviour of the pattern can be
  imitated through the composition of other patterns that are implemented by
  the WEE.

  \fitem{\rateC handler/external:} The pattern can be implemented in cooperation with
  an adopted handler wrapper or other external modules. These modules are still
  orchastrated by the WEE through the worfklow description but contribute the
  actual logic how to implement the pattern in coordination with the WEE.

  \fitem{\rateD orchestrated instances:} The pattern can only be expressed by
  controlling multiple instances of the execution engine. In addition, these
  instances have to be orchestrated and managed by the controller of the
  instances (e.g. the controller is responsible for data exchange between the
  instances). As the controller has the control about each instance, the
  orchestration can be itemized down to the execution of single activities. Of
  course, the pattern is then not implemented by the WEE, the logic is mainly
  within the controller.

\end{itemize}

\subsection{Basic Control Flow Patterns}

Each of the basic control flow patterns is directly supported by an explicit
WEE-DSL element. 

\begin{itemize}

  \fitem{\rateA Sequence}. The sequence pattern indicates that an activity is started after
  another activity finishes. An example implementation can be seen in Listing
  \ref{lst:basic_sequence}.

  \lstset{language=Ruby,caption=\textsc{Sequence pattern},label=lst:basic_sequence}
\begin{lstlisting}
activity :a1, :call, endpoint1
activity :a2, :call, endpoint2
\end{lstlisting}

  \fitem{\rateA \rateA  Parallel split \& Synchronization}. The parallel split generates
  multiple branches which have to be executed concurrently. The synchronization
  merges this branches together as soon as all parallel branches have finished.
  Listing \ref{lst:basic_parallel} shows the implementation in the WEE-DSL. 

  \lstset{language=Ruby,caption=\textsc{Parallel split \& Synchronization},label=lst:basic_parallel}
\begin{lstlisting}
parallel :wait do
  parallel_branch do
    activity :branch1_a1, :call, endpoint1
  end
  parallel_branch do
    activity :branch2_a1, :call, endpoint2
  end
end
\end{lstlisting}

  \fitem{\rateA \rateA  Exclusive Choice \& Simple Merge}. Exclusive choice selects one of
  multiple possible branches to be executed. Which branch to continue on is
  decided by a condition. The simple merge joins multiple branches into a
  single subsequent branch. The subsequent branch is executed as soon as one of
  the multiple branches has finished. The implementation of these two patterns
  can be seen in Listing \ref{lst:basic_choice}. Multiple branches can be
  defined by the keyword \textit{alternative}. Each alternative is guarded with
  a condition that indicates in which situation the branch should be executed
  as indicated in line \ref{choice:alternative}.  If none of the defined
  alternatives is executed, the optional \textit{otherwise}-section is executed
  as indicated in line \ref{choice:otherwise}.

  \lstset{language=Ruby,caption=\textsc{Exclusive choice \& Simple Merge},label=lst:basic_choice}
\begin{lstlisting}
context :x => 5
choose do
  alternative(@x > 1) do               (*@\label{choice:alternative}@*)
    activity :alternative1_a1, :call, endpoint1
  end
  otherwise do                         (*@\label{choice:otherwise}@*)
    activity :alternative2_a1, :call, endpoint2
  end
end
\end{lstlisting}

\end{itemize}

\subsection{Advanced Branching and Synchronization Patterns} 

\begin{itemize}

  \fitem{\rateA \rateD \rateA Multi-Choice, Multi-Merge \& Structured Synchronizing Merge}. The
  multi-choice pattern extends the exclusive choice pattern by allowing
  multiple branches to be executed. The merge of these branches therefore has
  to merge one or more possible branches. The WEE-DSL does not distinguish
  between these patterns and the exclusive choice \& the simple merge pattern.
  An example of a multi-choice with two executed branches can be seen in
  listing \ref{lst:adv_multi}. The alternatives in line
  \ref{multi:alternative1} and line \ref{multi:alternative2} are both executed
  in this example. In this case the branches are not executed concurrent but
  one after another. Listing \ref{lst:adv_multi_concurrent} shows an example
  how to implement the patterns to execute the branches concurrent which
  results in a structured synchronizing merge. Each
  \textit{parallel\_branch}-block is executed in a separated thread. To
  indicate which threads are running parallel, a \textit{parallel}-block has to
  enclose the \textit{choice}-block.  The multi-merge pattern distinguishes
  from the structured synchronizing merge by keeping the multiple threads of
  control beyond the merging point.  If the multi choice pattern results in the
  execution of two or more branches, each thread of control is passed to the
  subsequent branch in the way that the subsequent branch is executed two or
  more times. The multi merge pattern is not supported by the WEE.

  \lstset{language=Ruby,caption=\textsc{Multi-Choice \& Structured Synchronizing Merge - Not concurrent},label=lst:adv_multi}
\begin{lstlisting}
context :x => 5
choose do
  alternative(@x > 1) do               (*@\label{multi:alternative1}@*)
    activity :alternative1_a1, :call, endpoint1
  end
  alternative(@x < 10) do               (*@\label{multi:alternative2}@*)
    activity :alternative2_a1, :call, endpoint2
  end
  otherwise do                         (*@\label{multi:otherwise}@*)
    activity :alternative3_a1, :call, endpoint3
  end
end
\end{lstlisting}

  \lstset{language=Ruby,caption=\textsc{Multi-Choice \& Structured Synchronizing Merge - Concurrent},label=lst:adv_multi_concurrent}
\begin{lstlisting}
context :x => 5
parallel do
  choose do
    alternative(@x > 1) do
      parallel_branch do
        activity :alternative1_a1, :call, endpoint1
      end
    end
    alternative(@x < 10) do
      parallel_branch do
        activity :alternative2_a1, :call, endpoint2
      end
    end
    otherwise do
      parallel_branch do
        activity :alternative3_a1, :call, endpoint3
      end
    end
  end
end

\end{lstlisting}

  \fitem{\rateD \rateD \rateA Structured Discriminator}. The structured discriminator pattern is
  similar to the structured synchronizing merge pattern. While the structured
  synchronizing merge pattern is used to join branches forked by a choice, the
  structured discriminator merges branches forked by a parallel split. Also,
  the structured discriminator does not wait until each branch finished but
  continues the subsequent branch as soon as \textit{one} branch finishes. Other
  parallel branches may be still executed and are terminated as they approach
  at the structured discriminator. The structured discriminator is not
  supported directly in the WEE-DSL. However, the structured discriminator has
  two variants.  
  
  The \textit{Blocking Discriminator pattern} does not allow to fork parallel
  branches while another instance is still executing one of the branches.  As
  the WEE focuses on the execution within one workflow instance, it does not
  support interactions or management between multiple instances. Therefore,
  this pattern is not supported.

  The \textit{Canceling Discriminator pattern} passes the thread of control to
  the subsequent branch as soon as one of the parallel branches has finished.
  All remaining parallel branches that are still executed are canceled. The
  WEE-DSL supports this pattern by providing a parameter to the parallel-block
  as can be seen in listing \ref{lst:adv_canceling_disc}.  The
  \textit{wait}-parameter can be used to set how much branches have to be
  finished before the execution in the subsequent branch is continued. The
  \textit{no-longer-necessary}-signal is sent to still running branches, urging
  them to quit execution.

  \lstset{language=Ruby,caption=\textsc{Canceling Discriminator},label=lst:adv_canceling_disc}
\begin{lstlisting}
parallel :wait => 1 do
  parallel_branch do
    activity :branch1_a1, :call, endpoint1
  end
  parallel_branch do
    activity :branch2_a2, :call, endpoint2
  end
end
\end{lstlisting}

  \fitem{\rateD \rateD \rateA Structured Partial Join}. The structured partial join pattern is
  similar to the structured discriminator but continues the subsequent branch
  as soon as a \textit{specified amount} of branches finishes execution.  As
  the WEE-DSL uses the same constructs for the Discriminator, the structural
  partial join and its variant, the \textit{Blocking Partial Join} are not
  supported. The \textit{Canceling Partial Join} works as a generalized version
  of the canceling discriminator, therefore the listing
  \ref{lst:adv_canceling_disc} shows both patterns.

  \fitem{\rateD Generalised AND-Join}. The generalised AND-join pattern merges
  branches as soon as all incoming branches have been completed.  In contrast
  to the synchronization pattern, the generalised AND-join supports the join
  across multiple threads of controls and passes the thread of control to the
  subsequent branch each time all incoming branches are completed. This is also
  true even when the parallel branches are not created by the same parallel
  split event.  The WEE does not support the generalised AND-join pattern as it
  is only able to merge branches that result from the same parallel split
  event.

  \fitem{\rateD \rateD Local Synchronizing Merge \& General Synchronizing Merge}. These two
  patterns can be used instead of the synchronization if the process
  description is not structured.  Both patterns are able to pass the thread of
  control to the subsequent branch if all branches are executed or if it is
  sure that all outstanding branches are not going to arrive at the merge
  point. Other than in the local synchronizing merge situation, the information
  if branches are going to arrive at the merge point is not available locally
  in the general synchronizing merge situation.  This information may be
  gathered by the evaluation of possible prospective states of the executed
  branch.  Both patterns cannot be translated into the WEE-DSL as the process
  definition is expressed in a block oriented manner. Therefore, the process
  description is available in a structured way which conflicts with the basic
  reason for the two patterns. 
  
  \fitem{\rateA \rateA Thread Split \& Thread Merge}. The thread split pattern generates a
  specific amount of threads. Each of the threads executes the subsequent
  branch. The thread merge pattern is able to merge multiple thread that are
  generated by the thread split pattern. The pattern description states that
  the number of threads that have to be generated must be specified at
  design-time. Although the concepts of threads within a workflow instance was
  not intended in the design of the WEE-DSL apart from parallel branches, the
  dynamic generation of threads can be achieved by cascading the
  \textit{parallel\_branch}-element into a loop (\textit{cycle}-element). As
  each \textit{parallel\_branch} forks a thread, this arrangement implements
  the thread-split and the thread-merge pattern through a minor modification of
  the workflow. An example implementation can be seen in listing
  \ref{lst:adv_thread}.

  \lstset{language=Ruby,caption=\textsc{Thread-Split \& Thread-Merge},label=lst:adv_thread}
\begin{lstlisting}
parallel :wait do
  context :x => 3                       (*@\label{thread:context}@*)
  cycle("@x >= 0") do
    activity :countdown, :manipulate do
      @x -= 1
    end
    parallel_branch do
      activity :a, :call, endpoint1     (*@\label{thread:activity}@*)
    end
  end
end
\end{lstlisting}

\end{itemize}

\subsection{State Based Patterns}

\begin{itemize}

  \fitem{\rateB Deferred Choice}. The deferred choice pattern chooses one of multiple
  possible branches to be executed. The decision, which branch has to be
  executed, is delayed as long as possible. In contrast to the choice-patterns,
  the decision is not made explicit by a condition. After one branch starts the
  execution, each other branch is aborted, therefore the decision is more based
  on a race of the branches which is executed next. There is no explicit
  element in the WEE-DSL to express the deferred choice pattern. However, with
  the implementation of the canceling discriminator, it is possible to achieve
  a similar result. A possible implementation can be seen in listing
  \ref{lst:state_deferred}. Each branch is represented by a single activity
  which is part of a \textit{parallel}-block (which in this case is a canceling
  discriminator). Line \ref{deferred:represent1} and line
  \ref{deferred:represent2} are showing the two representative activities. As
  soon as one of them completes, a context variable expressing which branch to
  execute is set (see line \ref{deferred:rep_set1} and line
  \ref{deferred:rep_set2}). As the canceling discriminator aborts all other
  branches, the context variable now indicates which branch has to be executed
  further. The use of the exclusive choice pattern is now possible (see line
  \ref{deferred:choose}).

  \lstset{language=Ruby,caption=\textsc{Deferred Choice},label=lst:state_deferred}
\begin{lstlisting}
context :choice => nil
parallel :wait => 1 do
  parallel_branch do
    activity :representA, :call, endpoint1, 1 do   (*@\label{deferred:represent1}@*)
      @choice = 1                                  (*@\label{deferred:rep_set1}@*)
    end
  end
  parallel_branch do
    activity :representB, :call, endpoint2, 2 do   (*@\label{deferred:represent2}@*) 
      @choice = 2                                  (*@\label{deferred:rep_set2}@*)
    end
  end
end
choose do                                          (*@\label{deferred:choose}@*)
  alternative(@choice == 1) do
    # branch 1 comes here
  end
  alternative(@choice == 2) do
    # branch 2 comes here
  end
end
\end{lstlisting}

  \fitem{\rateB Milestone}. The execution of an activity is enabled as long as a
  specific point in the execution of the workflow is reached. In general, the
  milestone pattern needs at least two parallel branches. One branch marks if
  the milestone is reached and enables the execution of specific activities in
  the other branch(es). These activities cannot be executed before the
  milestone is reached or after the milestone is passed. The WEE does not
  provide a direct implementation of the milestone pattern. In general, two
  approaches can be used two tackle the problem: 

  \begin{enumerate}

    \item The milestone is activated by an explicit activity and deactivated by
    a explicit activity. Before an activity which is enabled by the milestone
    is executed, a \textit{choose}-element checks for the activation. Listing
    \ref{lst:state_milestone} shows an example implementation. After activity
    \textit{:activate\_milestone} is executed (line \ref{milestone:activate}), the
    milestone is reached, which is indicated by the context variable
    \textit{milestone}. In the parallel branch, the activity \textit{:enabled} (line
    \ref{milestone:enabled}) is executed if the context variable
    \textit{milestone} is true when the \textit{choose}-element is reached
    (line \ref{milestone:choose}).  The milestone is valid as long as the
    activity \textit{:keep\_milestone} is running (line \ref{milestone:keep}). The
    activity \textit{:deactivate\_milestone} (line \ref{milestone:deactivate}) resets
    the context variable \textit{milestone} and the \textit{:enabled}-activity is not
    executed anymore. Instead, the \textit{:not\_enabled}-activity is called (line
    \ref{milestone:not_enabled}).

    \item The controller of the workflow ``injects" the enabled activity by
    restructuring the workflow. The workflow execution is suspended and the
    workflow description is adjusted by the enabled activity. When the
    milestone is expired, the workflow description is set back to the original.
    While doing this, the controller has to take care about all active
    activities and track the thread of control for each branch. In this
    scenario, the controller supervises when and how long a milestone becomes
    active. This alternative is more difficult to implement but applicable if
    the milestone logic should not be part of the workflow description. This
    may be true if specific activities should only be executed when certain
    technical issues are fulfilled (e.g. services are only temporary available
    or their call should be avoided normally)

  \end{enumerate}

  \lstset{language=Ruby,caption=\textsc{Milestone},label=lst:state_milestone}
\begin{lstlisting}
context :milestone => false
parallel do
  parallel_branch do
    activity :activate_milestone, :manipulate do  (*@\label{milestone:activate}@*)
      @milestone = true
    end
    activity :keep_milestone, :manipulate do      (*@\label{milestone:keep}@*)
      sleep 5
    end
    activity :deactivate_milestone, :manipulate do (*@\label{milestone:deactivate}@*)
      @milestone = false
    end
  end
  parallel_branch do
    choose do                                      (*@\label{milestone:choose}@*)
      alternative(@milestone) do
        activity :enabled, :call, endpoint1         (*@\label{milestone:enabled}@*)
      end
      otherwise do
        activity :not_enabled, :call, endpoint2    (*@\label{milestone:not_enabled}@*)
      end
    end
  end
end
\end{lstlisting}

  \fitem{\rateA Critical Section}. Two or more areas of different (parallel) branches
  are defined to be prohibited to be executed at the same time for a given
  workflow instance. When an activity of a critical section is executed, that
  critical section has to be completed before the critical section can be
  entered again. This pattern is also commonly described as mutex or semaphore.
  The WEE-DSL has a separate element to define critical sections. Listen
  \ref{lst:state_critical} shows an example of a critical section which spans
  over two parallel branches. As a critical section can span across different
  branches, an identifier is necessary to indicate the connection.

  \lstset{language=Ruby,caption=\textsc{Critical Section},label=lst:state_critical}
\begin{lstlisting}
parallel do
  parallel_branch do
    critical(:id) do
      activity :parallel1, :manipulate do
        sleep 2
      end
    end
  end
  parallel_branch do
    critical(:id) do
      activity :parallel2, :manipulate do
        sleep 2
      end
    end
  end
end

\end{lstlisting}

  \fitem{\rateA Interleaved Routing}. A set of branches has to be executed. The
  branches can be executed in any order but must not be executed concurrently.
  None of the branches must be active as long as another branch is executed.
  The thread of control is passed to the subsequent branch when all branches
  (that are part of the interleaved routing) are finished.  The WEE-DSL does
  not have an explicit element for the interleaved routing pattern. This
  pattern can be easily implemented by the use of the critical section pattern.
  Listing \ref{lst:state_interleaved} shows a sample implementation.  Two
  (initially parallel) branches are opened in line \ref{interleaved:branch1}
  and line \ref{interleaved:branch2}. As the branches share a critical section,
  they cannot run simultaneously but one after another.

  \lstset{language=Ruby,caption=\textsc{Interleaved Routing},label=lst:state_interleaved}
\begin{lstlisting}
parallel do
  parallel_branch do    (*@\label{interleaved:branch1}@*)
    critical(:id) do
      activity :branch1_task1, :call, endpoint1
      activity :branch1_task2, :call, endpoint2
    end
  end
  parallel_branch do    (*@\label{interleaved:branch2}@*)
    critical(:id) do
      activity :branch2_task1, :call, endpoint3
    end
  end
end

\end{lstlisting}

  \fitem{\rateA Interleaved Parallel Routing}. The interleaved parallel routing
  pattern enhances the interleaved routing pattern by restricting the execution
  to the activity-level. Parallel branches may be active but only one activity
  over all branches in the interleaved parallel routing is executed at one
  time. Again, this can be achieved by the use of the critical section pattern.
  Not the critical section has to allow a ``switch" in the execution of the
  parallel branches. Therefore, the \textit{critical}-elements have to span
  around each activity to give activities from other branches the chance to be
  executed. This can be seen in listing \ref{lst:state_interleaved_parallel}.

  \lstset{language=Ruby,caption=\textsc{Interleaved Parallel Routing},label=lst:state_interleaved_parallel}
\begin{lstlisting}
parallel do
  parallel_branch do
    critical(:id) do
      activity :branch1_task1, :call, endpoint1
    end
    critical(:id) do
      activity :branch1_task2, :call, endpoint2
    end
  end
  parallel_branch do
    critical(:id) do
      activity :branch2_task1, :call, endpoint3
    end
  end
end
\end{lstlisting}

\end{itemize}

\subsection{Multiple Instances} 

The multiple instances patterns are created to deal with the generation of
multiple instances of an activity or task within a given workflow instance. The
different patterns focus on aspects like the amount of created instances,
synchronization and merging. The generated instances run independent and
concurrent to each other but mostly need to be synchronized with the workflow
instance when finished. The WEE basically has no concept of multiple instances
for an activity. However, the multiple instance patterns overlaps with the
thread split/thread merge pattern and can be expressed by them. If this is not
the case, the behavior often can be simulated by a suitable handler wrapper
implementation. As the handler wrapper is ordered by the WEE with the execution
of an activity, the handler wrapper can spawn the needed amount of instances
and is able to control them as long as needed.

\begin{itemize}

  \fitem{\rateC Multiple Instances without Synchronization}. Multiple instances of an
  activity are created but run independent from the workflow instance. The
  subsequent branch does not have to wait for the execution of the multiple
  instances. The amount of instances is not defined, consequentially the
  timespan in which new instances can be created is not defined.  This multiple
  instance pattern cannot be expressed by the WEE-DSL. An adjusted handler
  wrapper can simulate the behavior by spawning the needed amount of instances
  in separated threads without giving the WEE notice. This solution does not
  address the problem if the workflow instance finishes before all instances
  spawned by the handler wrapper have finished.

  \fitem{\rateA Multiple Instances with a Priori Design-Time Knowledge}. A number of
  instances is created. The amount is specified at design time. The thread of
  control is passed to the subsequent path as soon as all instances have been
  finished. The WEE supports this pattern in the same way as the thread-split
  and thread-merge pattern. Listing \ref{lst:adv_thread} therefore can also be seen
  as an implementation of this pattern. The context variable \textit{x} (defined in line
  \ref{thread:context} determines the amount of created instances of the
  activity defined in line \ref{thread:activity}.

  \fitem{\rateA Multiple Instances with a Priori Run-Time Knowledge}. A number of
  instances is created. The amount of instances is determined by the workflow
  logic before the first creation of an instance. The WEE supports this pattern
  in the same way as the thread-split and thread merge pattern. In contrast to
  the solution above, the number of created instances is determined by an
  activity. Listing \ref{lst:mi_runtime} shows an example implementation of the
  pattern. In line \ref{mi:context}, the context variable is initialized to a
  default value. The activity \textit{determine} (line \ref{mi:determine1} to
  line \ref{mi:determine2}) sets the number of instances that have to be
  created. The loop (line \ref{mi:cycle}) then generates the multiple instances
  of activity \textit{a} (line \ref{mi:activity}).

  \lstset{language=Ruby,caption=\textsc{Multiple Instances with a Priori Run-Time Knowledge},label=lst:mi_runtime}
\begin{lstlisting}
parallel :wait do
  context :x => 0                       (*@\label{mi:context}@*)
  activity :determine, :call, endpoint1 do |count|  (*@\label{mi:determine1}@*)
    @x = count                                    
  end                                               (*@\label{mi:determine2}@*)
  cycle("@x >= 0") do                               (*@\label{mi:cycle}@*)
    activity :countdown, :manipulate do
      @x -= 1
    end
    parallel_branch do
      activity :a, :call, endpoint1     (*@\label{mi:activity}@*)
    end
  end
end
\end{lstlisting}

  \fitem{\rateA Multiple Instances without a Priori Run-Time Knowledge}.  A number of
  instances have to be created. The number of instances is not determined until
  the last instance has finished execution.  The logic if more instances have
  to be created cannot be determined a priori but is derived from the
  execution process, resources or external services. As the logic of creating
  further instances has to be placed inside the workflow description, the
  implementation differs from the multiple instances patterns above. The
  execution of the subsequent branch has to be blocked until the last instance
  has finished the execution.  Listing \ref{lst:mi_without_runtime} shows an
  example implementation with the WEE-DSL.  In line \ref{mi_without:context} a
  context variable is defined which indicates if a new instance of the activity
  has to be created. This is at least true for the first time. The loop in line
  \ref{mi_without:cycle} creates a new instance of activity \textit{a} (line
  \ref{mi_without:activity}) as long as the context variable
  \textit{create\_instance} is true. The context variable
  \textit{create\_instance} is set by an external service (line
  \ref{mi_without:decide}) which defines if another instance of activity
  \textit{a} has to be created. As the call of the external service is not part
  of the parallel execution, it is independent from the execution of the
  activity instance. Therefore it can determine the creation time for each
  activity instance. Multiple instances can be created parallel or one after
  another by blocking the service call until an instance has finished
  execution. 

  \lstset{language=Ruby,caption=\textsc{Multiple Instances without a Priori Run-Time Knowledge},label=lst:mi_without_runtime}
\begin{lstlisting}
parallel :wait do
  context :create_instance => true                    (*@\label{mi_without:context}@*)
  cycle("@create_instance") do                        (*@\label{mi_without:cycle}@*)
    parallel_branch do
      activity :a, :call, endpoint1                   (*@\label{mi_without:activity}@*)
    end
    activity :decide, :call, endpoint2 do |result|      (*@\label{mi_without:decide}@*)
      @create_instance = result
    end
  end
end
\end{lstlisting}

  \fitem{\rateD \rateA \rateD Static Partial Join for Multiple Instances}. The static partial join
  for multiple instances forwards the thread of control to the subsequent
  branch as soon as a given amount of instances finished execution. Other,
  still running instances, have to be completed to re-enable the pattern but
  the result of these instances is withdrawn. The WEE does not support this
  pattern as it can only join all finished instances or abort instances which
  are no longer necessary. There are two variants of this pattern: The
  \textit{Canceling Partial Join for Multiple Instances}  pattern is analogue
  to the canceling partial join pattern described above. Other than the merge
  of branches, the canceling partial join for multiple instances pattern joins
  a given number of instances. The number of instances can be determined at
  design time or before the first instance is created.  Listing
  \ref{lst:mi_canceling} shows how to implement this pattern with the WEE-DSL.
  Multiple \textit{parallel-branch}es (and therefore multiple instances of
  activity \textit{a}) are created by the loop (line
  \ref{mi_canceling:cycle_start} to \ref{mi_canceling:cycle_end}). As each
  \textit{parallel\_branch} belong to the \textit{parallel}-block defined in
  line \ref{mi_canceling:parallel}, the number of \textit{parallel\_branches}
  (and therefore the number of instances) that have to be completed can be
  defined here. The \textit{parallel}-construct forwards the thread of control
  to the subsequent path as soon as the given number of
  \textit{parallel\_branches} has completed. The
  \textit{nolongernecessary}-signal is sent to unfinished
  \textit{parallel\_branches}.

  \lstset{language=Ruby,caption=\textsc{Canceling Partial Join for Multiple Instances},label=lst:mi_canceling}
\begin{lstlisting}
parallel :wait => 1 do                  (*@\label{mi_canceling:parallel}@*)
  context :x => 3                       (*@\label{mi_canceling:context}@*)
  cycle("@x >= 0") do                   (*@\label{mi_canceling:cycle_start}@*)
    activity :countdown, :manipulate do
      @x -= 1
    end
    parallel_branch do
      activity :a, :call, endpoint1     (*@\label{mi_canceling:activity}@*)
    end
  end                                   (*@\label{mi_canceling:cycle_end}@*)
end
\end{lstlisting}

  The \textit{Dynamic Partial Join for Multiple Instances} pattern provides the
  most flexibility when dealing with the merge of multiple instances. New
  instances can be created as long as the last instance has not finished. The
  number of instances depends on the execution progress or external
  information. After an instance has been executed, the thread of control can
  be passed to the subsequent branch. All other running instances are then
  withdrawn. The WEE does not support this pattern as the number of instances
  which have to be completed must be known at the first instance creation.

\end{itemize}

\subsection{Cancellation and Force Completion}

\begin{figure}
  \begin{center}
    \includegraphics[width=0.8\textwidth]{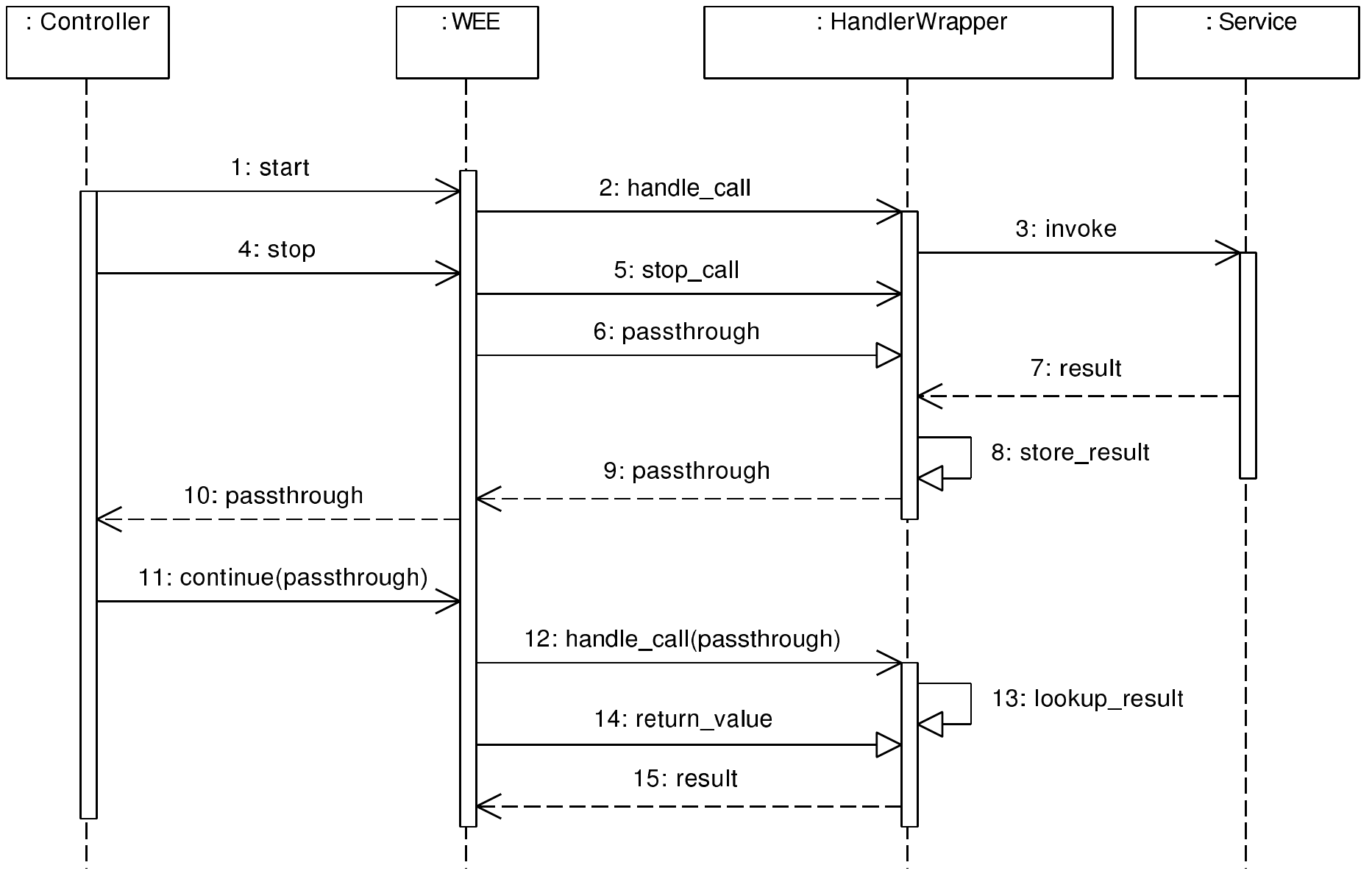}
  \end{center}
  \caption{Cancel Region Pattern}
  \label{fig:cancel_region}
\end{figure}

\begin{itemize}

  \fitem{\rateA \rateA Cancel Task \& Cancel Case}. The workflow instance is aborted and
  removed from execution. The cancel task pattern and the cancel case pattern
  are distinguished by the option \textit{when} the cancellation is possible:
  The cancel task pattern specifies that the cancellation is possible at a
  specific activity. This activity and further also the workflow instance is
  aborted. The cancel case pattern allows to abort a workflow instance not only
  at the point of the execution of an activity but for the whole execution of
  the instance.  The WEE does not distinguish between these patterns. It allows
  to stop the execution at any point during execution. The controller of the
  workflow instance has the possibility to send the stop-signal at any time. If
  a \textit{call}-activity is executed, the
  execution is delegated to the handler wrapper. Therefore the WEE cannot
  guarantee that the activity is aborted. The WEE can only inform the handler
  wrapper that it should stop the execution. The handler wrapper is then
  responsible to decide whether the service call can be aborted or not. This is
  important as the immediate abortion may lead to an inconsistent state or
  cannot be done without sanction which has to be avoided by the handler
  wrapper. 

  \fitem{\rateC Cancel Region}. Other than the cancel task and cancel case pattern,
  the cancel region pattern may not lead into the cancellation of the workflow
  instance.  If an active branch (or parts of it) is outside a cancel region,
  this branch is unaffected by the cancellation. The cancel region pattern
  defines a set of activities (which may be located also in different
  branches). Activities which are in the process of execution within this area
  are aborted when the cancel region pattern is activated. Activities which are
  in the process of execution outside of this area stay unaffected.  To
  implement this pattern with the WEE comes with some difficulties.  The WEE
  cannot abort only parts of the execution. If the WEE receives the stop-signal
  from the controller, the workflow instance is stopped. Therefore, also
  activities which should stay in the process of execution are affected, which
  is not the intention of the pattern. Even if the handler wrapper does not
  allow the abortion, the result of the service call will not be integrated
  into the workflow by the WEE. The handler wrapper will not even be asked to
  provide the result value. But the WEE will ask for a \textit{passthrough}
  value which gives the handler wrapper the possibility to store the result and
  reuse it when the workflow is continued. The sequence diagram in Fig.
  \ref{fig:cancel_region} now shows how the cancel region pattern can be
  implemented by the WEE. After the handler wrapper receives the
  \textit{stop\_call}-signal, the service call is continued unaffected and the
  result of the service call is stored. The passthrough-value (identifying the
  result for later use) is returned to the controller of the workflow instance.
  When the controller continues the execution, the thread of control is set to
  the activities outside of the cancel region. The handler wrapper is provided
  with the passthrough-value and can look up the stored result. Therefore the
  service call does not have to be repeated (which may be costly).

  \fitem{\rateA Cancel Multiple Instance Activity}. Multiple instances of an activity
  are executed within one workflow instance. The cancel multiple instance
  activity pattern aborts the workflow instance and subsequentially all active
  instances of an activity. Already completed instances of an activity are
  unaffected. The WEE covers this pattern with the same procedure as the cancel
  case pattern. The WEE allows to stop a workflow instance at any point during
  execution, therefore it is also possible to abort running instances of an
  activity.

  \fitem{\rateD Complete Multiple Instance Activity}. Multiple instances of an
  activity are created. During the execution of the instances, the pattern
  indicates that the subsequent branch has to be executed. Therefore, remaining
  instances are no longer executed and withdrawn. Already finished instances
  are synchronized and their results are integrated into the workflow. The
  thread of control is passed to the subsequent branch.  The WEE does not
  support this pattern. Although the WEE supports the manipulation of the
  thread of control, it is necessary to stop the execution before the
  manipulation can be done, which is not in the intention of the pattern.

\end{itemize}

\subsection{Termination and Triggers}

\begin{itemize}

  \fitem{\rateA Implicit Termination}. A branch terminates if no further control
  structures have to be executed. If all branches have terminate, the workflow
  instance is also terminated and the workflow execution is marked as completed
  successfully. Due to the block structure of the WEE-DSL, the implicit
  termination is inherent. 

  \fitem{\rateA Explicit Termination}. The workflow instance is terminated as soon as
  a specified point in the execution is reached. All remaining branches are
  canceled. The WEE allows the handler wrapper or the controller of the
  workflow to stop the execution at any point during the execution by sending
  the \textit{stop}-signal. The same mechanism is provided to the workflow
  description. Listing \ref{lst:termination_explicit} shows an example
  implementation. The activity in line \ref{explicit:stop} sends the
  \textit{stop}-signal to the WEE. If the handler wrapper has still running
  service calls, the handler wrapper is ordered to stop them if possible. Other
  activities will not be started. In contrast to the implicit termination, the
  end state of the workflow instance is set to stopped. 

  \lstset{language=Ruby,caption=\textsc{Explicit Termination},label=lst:termination_explicit}
\begin{lstlisting}
activity :stop, :manipulate do   (*@\label{explicit:stop}@*)
  stop                  
end
activity :whatever, :manipulate do
  # will not be executed
end
\end{lstlisting}

  \fitem{\rateC \rateC Persistent \& Transient Trigger}. An activity is enabled by a trigger.
  The persistent trigger pattern is implemented if the trigger event is stored
  until the activity is scheduled for execution. The event enables the activity
  during any time in the workflow execution.  The transient trigger enables an
  activity only when it is scheduled for execution. If the transient trigger
  event occurs before the activity is scheduled for execution, the event is
  withdrawn. An activity which has not been enabled by a trigger blocks the
  execution until a trigger event occurs or the workflow execution is canceled.
  The WEE does not support triggers directly.  As the execution of a service
  call is delegated to the handler wrapper, the trigger logic can be placed
  there. The handler wrapper can block the execution until the trigger event
  occurs. The downside of this approach is that the triggering logic is not
  anymore in the workflow description.

\end{itemize}

\subsection{Iterations}

\begin{figure}
  \begin{center}
    \includegraphics[width=0.5\textwidth]{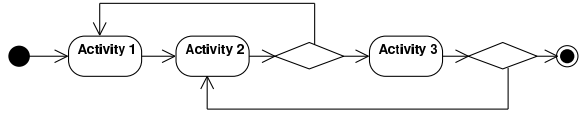}
  \end{center}
  \caption{Arbitrary Cycles Pattern Example}
  \label{fig:arbitrary_cycles_example}
\end{figure}

\begin{figure}
  \begin{center}
    \includegraphics[width=0.6\textwidth]{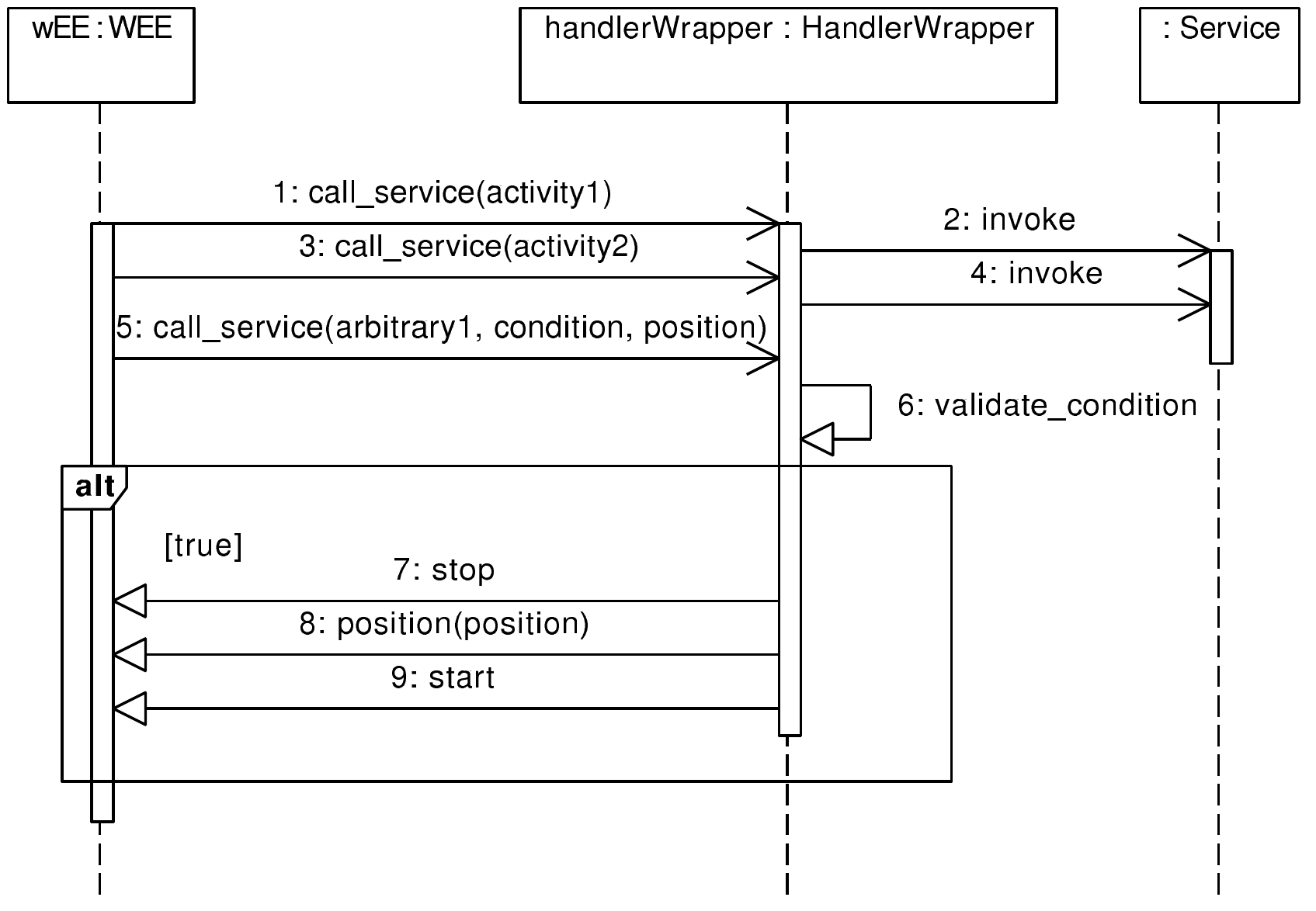}
  \end{center}
  \caption{Arbitrary Cycles Pattern Implementation}
  \label{fig:arbitrary_cycles}
\end{figure}

\begin{itemize}

  \fitem{\rateC Arbitrary Cycles}. The arbitrary cycles pattern allows unstructured
  loops. The thread of control can be set back to a defined point in the
  workflow description. The defined point can be located within another
  arbitrary cycle. Fig. \ref{fig:arbitrary_cycles_example} shows an example of
  a of an arbitrary cycle. As the WEE-DSL has a block structure, the arbitrary
  cycles pattern can not be expressed directly. Instead, the WEE allows
  modification of the thread of control which can be used to simulate the
  behavior. Fig. \ref{fig:arbitrary_cycles} shows how a modified handler
  wrapper has to act to implement the pattern. Activity 1 and Activity 2 are
  performed normally as service calls. The next \textit{call\_service}-message
  is then interpreted by the handler wrapper. The decision if the thread of
  control has to be altered is based on the condition provided by the workflow
  description. If the condition is validated as true, the thread of control is
  set to the given position identifier. The necessary information can therefore
  be specified in the workflow description and is not hard-wired in the handler
  wrapper. Although this implements the arbitrary cycles pattern, a complete
  implementation of the handler wrapper may also take care of possible
  side-effects which can arise when dealing with parallel active branches.

  \fitem{\rateA Structured Loop}. An activity or a set of activities and control
  structures are executed repetitive. The loop can be top-controlled or
  bottom-controlled. In contrast to the arbitrary cycles pattern, a structured
  loop can be expressed in a block manner. The WEE-DSL directly supports the
  structured loop pattern with the \textit{cycle}-element. This element
  implements the top-controlled loop. Bottom-controlled loops are not supported
  but can be expressed through a top-controlled loop with minor overhead.
  Listing \ref{lst:iteration_structured} shows a sample of a structured loop.

  \lstset{language=Ruby,caption=\textsc{Structured Loop},label=lst:iteration_structured}
\begin{lstlisting}
context :x => 3
cycle("@x > 0") do
  activity :countdown, :manipulate do
    @x -= 1
  end
end
\end{lstlisting}

  \fitem{\rateC Recursion}. The execution of an activity results in the
  execution of a new instance of the overall workflow description that is
  currently executed.  Each recursion has to have at least one exit condition.
  As the WEE does not has the concept of multiple instances, the recursion
  cannot be supported directly. Instead, a recursion can be seen as a normal
  service call which in turn invokes a new workflow instance. The
  implementation can be therefore done by the handler wrapper.

\end{itemize}

\subsection{Comparison to Other Engines}

When we compare WEE with other workflow engines, the boundaries of our
execution engine become apparent. Our execution engine focuses on the control
flow aspect of workflows. More precisely, we even focus on a single thread of
control within our execution engine (except parallel branches). 

Existing workflow engines on the other hand have to cover a much larger set of
functions and do not only focus on the control flow aspect. They also provide
support for data and resource handling, security, logging, repair strategies
and much more. 

Table \ref{tab:eval_summary} shows a summary of the pattern coverage as
described so far.

\vspace{3cm}

\renewcommand{\arraystretch}{1.3} 
\begin{center}
  \begin{longtable}{p{2.5cm}|l|c|c|c|c}
    \caption{Workflow Patterns Coverage} \\
    \label{tab:eval_summary} \\
    \hline
    \bfseries Pattern class & \bfseries Pattern name & \bfseries \rotatebox{90}{directly supported} & \bfseries \rotatebox{90}{modified workflow} & \bfseries \rotatebox{90}{handler/external} & \bfseries \rotatebox{90}{\minibox{orchestrated\\ instances}} \\
    \hline\hline
    \endfirsthead
    \hline
    \bfseries Pattern class & \bfseries Pattern name & \bfseries \rotatebox{90}{directly supported} & \bfseries \rotatebox{90}{modified workflow} & \bfseries \rotatebox{90}{handler/external} & \bfseries \rotatebox{90}{\minibox{orchestrated\\ instances}} \\
    \hline\hline
    \endhead

    \multirow{5}{*}{\rotatebox{90}{\minibox{Basic Control\\ Flow}}}    
                          & Sequence          & \Heart\Heart &  & &	\\ \cline{2-6}
                          & Parallel Split    & \Heart\Heart &  & &	\\ \cline{2-6}
                          & Synchronization   & \Heart\Heart &  & &	\\ \cline{2-6}
                          & Exclusive Choice  & \Heart\Heart &  & &	\\ \cline{2-6}
                          & Simple Merge      & \Heart\Heart &  & &	\\
    \hline\hline
    \multirow{14}{*}{\rotatebox{90}{\minibox{Advances Branching and Synchronization}}} 
                          & Multi-Choice                   & \Heart\Heart &  & &	\\ \cline{2-6}
                          & Structured Synchronizing Merge & \Heart\Heart &  & &	\\ \cline{2-6}
                          & Multi-Merge                    &   &  & & \texttimes \\ \cline{2-6}
                          & Structured Discriminator       &   &  & &	\texttimes \\ \cline{2-6}
                          & Blocking Discriminator         &   &  & &	\texttimes \\ \cline{2-6}
                          & Cancelling Discriminator       & \Heart\Heart &  & &	\\ \cline{2-6}
                          & Structured Partial Join        &   &  & &	\texttimes \\ \cline{2-6}
                          & Blocking Partial Join          &   &  & &	\texttimes \\ \cline{2-6}
                          & Cancelling Partial Join        & \Heart\Heart &  & &	\\ \cline{2-6}
                          & Generalised AND-Join           &   &  & &	\texttimes \\ \cline{2-6}
                          & Local Synchronizing Merge      &   &  & &	\texttimes \\ \cline{2-6}
                          & General Synchronizing Merge    &   &  & &	\texttimes \\ \cline{2-6}
                          & Thread Merge                   & \Heart\Heart &  & &	\\ \cline{2-6}
                          & Thread Split                   & \Heart\Heart &  & &	\\
    \hline\hline
    \multirow{7}{*}{\rotatebox{90}{\minibox{Multiple Instances}}} 
                          & \minibox{Multiple Instances without\\ Synchronization}             &   &   & \textasteriskcentered &   \\ \cline{2-6}
                          & \minibox{Multiple Instances with a\\ Priori Design-Time Knowledge} & \Heart\Heart &   &   &   \\ \cline{2-6}
                          & \minibox{Multiple Instances with a\\ Priori Run-Time Knowledge}    & \Heart\Heart &   &   &   \\ \cline{2-6}
                          & \minibox{Multiple Instances without\\ a Priori Run-Time Knowledge} & \Heart\Heart &   &   &   \\ \cline{2-6}
                          & \minibox{Static Partial Join for\\ Multiple Instances}             &   &   &   & \texttimes \\ \cline{2-6}
                          & \minibox{Cancelling Partial Join for\\ Multiple Instances}         & \Heart\Heart &   &   &   \\ \cline{2-6}
                          & \minibox{Dynamic Partial Join for\\ Multiple Instances}            &   &   &   & \texttimes \\
    \hline\hline
    \multirow{5}{*}{\rotatebox{0}{\minibox{State Based}}}    
                          & Deferred Choice               &   & \Heart &   &	  \\ \cline{2-6}
                          & Interleaved Parallel Routing  & \Heart\Heart &   &   &	  \\ \cline{2-6}
                          & Milestone                     &   & \Heart &   &	  \\ \cline{2-6}
                          & Critical Section              & \Heart\Heart &   &   &	  \\ \cline{2-6}
                          & Interleaved Routing           & \Heart\Heart &   &   &	  \\
    \hline\hline
    \multirow{5}{*}{\rotatebox{0}{\minibox{Cancellation\\ and Force\\ Completion}}}    
                          & Cancel Task                          & \Heart\Heart &   &   &   \\ \cline{2-6}
                          & Cancel Case                          & \Heart\Heart &   &   &   \\ \cline{2-6}
                          & Cancel Region                        &   &   & \textasteriskcentered &   \\ \cline{2-6}
                          & Cancel Multiple Instance Activity    & \Heart\Heart &   &   &   \\ \cline{2-6}
                          & Complete Multiple Instance Activity  &   &   &   & \texttimes \\
    \hline\hline
    \multirow{3}{*}{\rotatebox{0}{\minibox{Iteration}}}    
                          & Arbitrary Cycles       &   &  & \textasteriskcentered &	\\ \cline{2-6}
                          & Structured Loop        & \Heart\Heart &  &   &	\\ \cline{2-6}
                          & Recursion              &   &  & \textasteriskcentered &	\\
    \hline\hline
    \multirow{2}{*}{\rotatebox{0}{\minibox{Termination}}}    
                          & Implicit Termination   & \Heart\Heart &  & &	\\ \cline{2-6}
                          & Explicit Termination   & \Heart\Heart &  & &	\\
    \hline\hline
    \multirow{2}{*}{\rotatebox{0}{\minibox{Trigger}}}    
                          & Transient Trigger      &  &  & \textasteriskcentered &	\\ \cline{2-6}
                          & Persistent Trigger     &  &  & \textasteriskcentered &	\\
    \hline\hline
\end{longtable}
\end{center}

Wohed et al. \cite{wohed_patterns-based_2007} created a pattern based
evaluation of different open source workflow systems. The analysis of our
pattern support above allows to compare the coverage of our execution engine
with this evaluation. In contrast to Wohed et al., we used a more detailed
categorization with levels of support instead of just coverage. To allow a
comparison, we assume that \textit{directly supported} (\Heart\Heart) is equal to ``+'',
\textit{modified workflow} (\Heart) and \textit{handler/external} (\textasteriskcentered) are equal to ``+/-''
and \textit{orchestrated instances} (\texttimes) refers to ``-'' as this is the least
supported category.

\begin{table}
  \renewcommand{\arraystretch}{1.3}
  \caption{Support of Control Flow Patterns}
  \label{tab:comp_result}
  \centering
  \begin{tabular}{l|r|r|r}
    \hline
    \bfseries Product & \bfseries \textit{+} & \bfseries \textit{+/-} & \bfseries \textit{-} \\
    \hline\hline
    WEE                  & 22 & 10 & 11	\\
    \hline
    StaffWare 10         & 14 & 0 & 29	\\
    \hline
    WebSphere MQ 3.4     & 11 & 0 & 32	\\
    \hline
    Oracle BPEL PM 10.12 & 18 & 6 & 19	\\
    \hline
    JBoss jBPM 3.1.4.2   & 13 & 2 & 28	\\
    \hline
    OpenWFE 1.7.3        & 20 & 4 & 19	\\
    \hline
    Enhydra Shark 2.0    & 11 & 0 & 32	\\
    \hline
  \end{tabular}
\end{table}

The results of the comparison (see Tab. \ref{tab:comp_result}) show that WEE,
although minimal, covers a range of patterns that enables it to outperform
several commercial solutions. We conclude therefore that converting workflow
description languages to the WEE-DSL can be achieved with low effort.

  \section{Conclusion} 
\label{sec:conclusion}

In this technical report we outlined the structure of our DSL as well as evaluated
its the supplied vocabulary against existing patterns. Additionally we compared the
results to existing products.

We proved that our approach is well suited to serve as a bridging technology.
We envision that WEE is the core of a new series of network based execution
engines that take arbitrary workflow descriptions (transformed to the DSL) and
execute it providing superior transparency. We envision external modules for
monitoring, repair and semantic rule checking to become common network based
infrastructure that is shared process aware information systems.


  \bibliographystyle{ieeetr}  
  \bibliography{roemer,ours}
\end{document}